\documentclass[10pt]{article} 

\usepackage{amsmath}
\usepackage{amssymb}

\usepackage{graphicx}

\usepackage{cite}

\usepackage{color}
\usepackage{soul} 

\usepackage{setspace}
\usepackage[labelfont=bf,labelsep=period,justification=raggedright]{caption}

\usepackage{hyperref}

\makeatletter
\renewcommand{\@biblabel}[1]{\quad#1.}
\makeatother

\date{}

\usepackage{bm} 

\begin{document}

\begin{flushleft}
{\Large
\textbf{Poly-Omic Prediction of Complex Traits: OmicKriging}
}
\\
Heather E. Wheeler$^{1}$, 
Keston Aquino-Michaels$^{2}$,
Eric R. Gamazon$^{2}$, 
Vassily V. Trubetskoy$^{2}$,
M. Eileen Dolan$^{1}$,
R. Stephanie Huang$^{1}$,
Nancy J. Cox$^{2}$,
Hae Kyung Im$^{3,\ast}$
\\
\bf{1} Section of Hematology/Oncology, Department of Medicine, University of Chicago, Chicago, IL, USA
\\
\bf{2} Section of Genetic Medicine, Department of Medicine, University of Chicago, Chicago, IL, USA
\\
\bf{3} Department of Health Studies, University of Chicago, Chicago, IL, USA
\\
$\ast$ E-mail: haky@uchicago.edu
\end{flushleft}

\section*{Abstract}

High-confidence prediction of complex traits such as disease risk or drug response is an ultimate goal of personalized medicine. Although genome-wide association studies have discovered thousands of well-replicated polymorphisms associated with a broad spectrum of complex traits, the combined predictive power of these associations for any given trait is generally too low to be of clinical relevance. We propose a novel systems approach to complex trait prediction, which leverages and integrates similarity in genetic, transcriptomic  or other omics-level data. We translate the omic similarity into phenotypic similarity using a method called Kriging, commonly used in geostatistics and machine learning. Our method called OmicKriging emphasizes the use of a wide variety of systems-level data, such as those increasingly made available by comprehensive surveys of the genome, transcriptome and epigenome, for complex trait prediction. Furthermore, our OmicKriging framework allows easy integration of prior information on the function of subsets of omics-level data from heterogeneous sources without the sometimes heavy computational burden of Bayesian approaches. Using seven disease datasets from the Wellcome Trust Case Control Consortium (WTCCC), we show that OmicKriging allows simple integration of sparse and highly polygenic components yielding comparable performance at a fraction of the computing time of a recently published Bayesian sparse linear mixed model method. Using a cellular growth phenotype, we show that integrating mRNA and microRNA expression data substantially increases performance over either dataset alone. We also integrate genotype and expression data to predict change in LDL cholesterol levels after statin treatment and show that OmicKriging performs better than the polygenic score method. We provide an R package to implement OmicKriging\\ (\url{http://www.scandb.org/newinterface/tools/OmicKriging.html}). 


\section*{Author Summary}
Recent advances in the development of the technology to query the structure and function of DNA, RNA, protein and other biomolecules set the stage for the use of genomic information for individualized risk prediction and optimized clinical care. In this manuscript, we propose a novel systems approach to complex trait prediction. We translate the similarity in DNA, RNA levels or other biomolecular profiles into phenotypic similarity using a method called Kriging, commonly used in geostatistics and machine learning. Our method called OmicKriging emphasizes the integrative use of a wide variety of systems-level data, such as those increasingly made available by comprehensive surveys of the genome, transcriptome and epigenome, for complex trait prediction. We apply our method to several human complex traits and evaluate its performance. We show that combining mRNA and microRNA transcriptome data is highly predictive of cellular growth, while using genetic variant data alone is not predictive. We show that OmicKriging using genotype and expression data predicts a clinical statin response phenotype better than the polygenic score method. In a study of seven common diseases from the Wellcome Trust Case Control Consortium, we show performance of OmicKriging is similar to more computationally intensive approaches. 

\section*{Introduction}
High-confidence prediction of  complex traits is an ultimate goal of personalized medicine. For heritable traits, genotypic similarity contributes to phenotypic similarity. While genome-wide association studies (GWAS) have revealed many statistically significant loci associated with complex traits, the small effect sizes of most loci limit their prediction utility\cite{de2010predicting}. For most complex traits, an appreciable proportion of phenotypic variance is explained only when polygenic models, rather than single-marker tests, are used\cite{purcell2009common,yang2010common}. For example, 45\% of the variance in human height can be explained using a mixed linear modeling approach (GCTA) that simultaneously considers all 300K common SNPs genotyped\cite{yang2010common,yang2011gcta}. Because this approach does not rely on selection of loci, it is thought to be appropriate for traits with a highly polygenic or infinitesimal genetic architecture\cite{makowsky2011beyond,ober2012using}.

\begin{figure}[!ht]
\begin{center}
\includegraphics[width=6in]{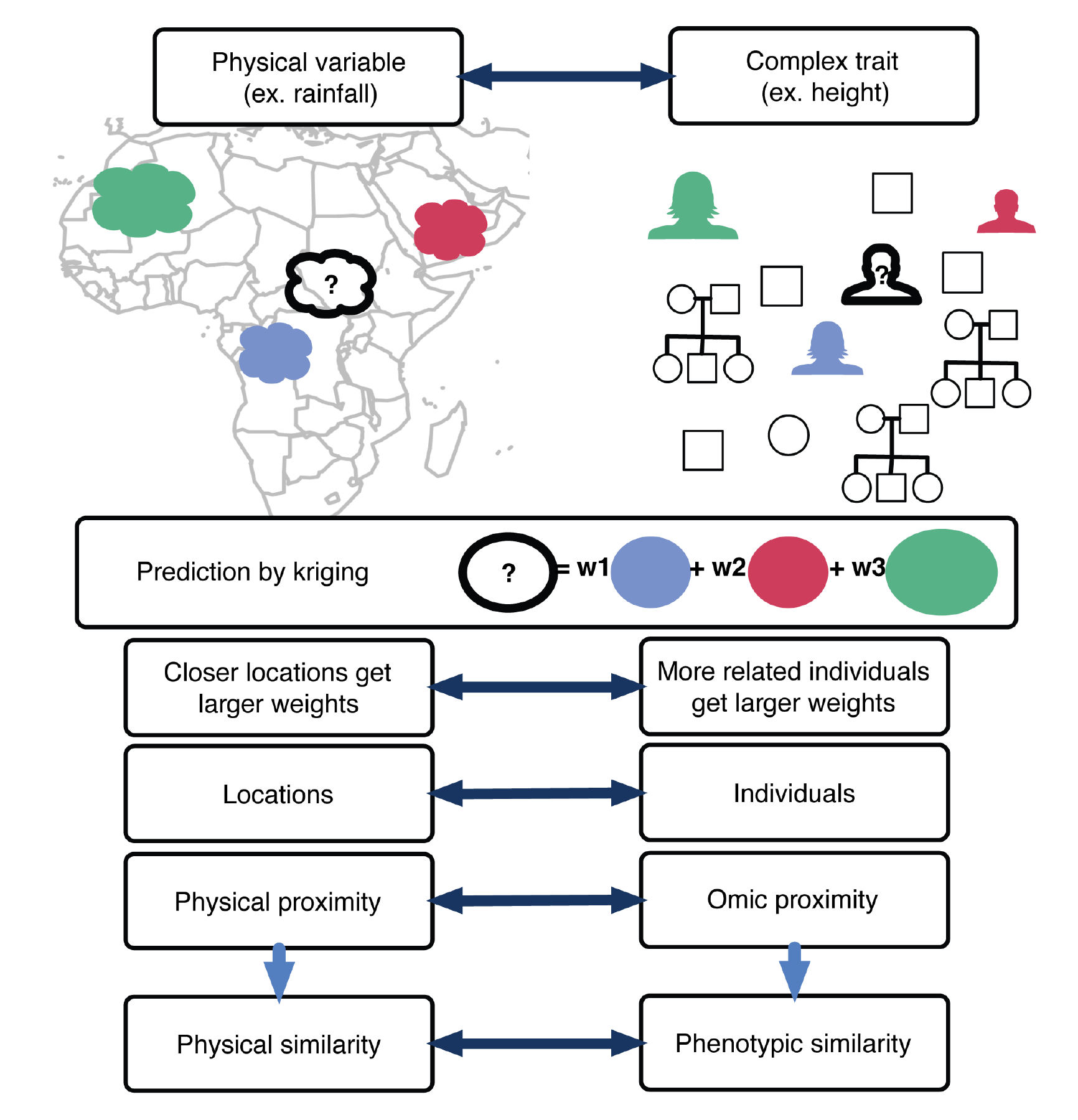} 
\end{center}
\caption{
{\bf Kriging and whole-genome prediction connection.}  This figure shows the analogous relationships between components of the kriging method used in geostatistics and whole-genome prediction. The prediction at an unobserved location (?) is computed as a weighted average of the variable at observed locations. The weights are functions of the correlation between the rainfall at the new location and the rainfalls at the observed locations. The closer the distance between each observed location and the new location, the higher the weight. In complex trait prediction, locations correspond to individuals, physical proximity corresponds to genetic relatedness. The correlation between two locations or individuals is the key component of this method. In animal breeding approaches, the genetic relatedness matrix or kinship matrix is used. In OmicKriging, a genetic relatedness matrix, a gene expression similarity matrix, or any combination of available high-throughput data similarity measures can be tested for complex trait prediction performance. 
}
\label{kriging-flow}
\end{figure}

In order to intuitively understand our approach to phenotype prediction from genomic data, an analogy from the field of geostatistics is useful. Kriging interpolates the value of some geographical measurement such as yearly rainfall at an unobserved location\cite{cressie1993statistics,stein1999interpolation,im2007semiparametric}. The approach assumes the rainfall measured at nearby locations will be more similar to that at the unmeasured location than rainfall at sites farther away. Kriging uses weights based on the correlation matrix of these locations along with the observed rainfall amounts to predict the rainfall at the unmeasured location. The locations are analogous to individuals and the rainfall amounts are analogous to the phenotype of interest (Figure \ref{kriging-flow}). In our application to complex trait prediction, the notion of distance can be replaced with the degree of genetic similarity or dissimilarity. Close ties between genetic distance and geographic distance have been demonstrated in studies of human population structure. For example, in the analysis of genome-wide genotype data from 3000 Europeans, a geographic map of Europe arose naturally when the first two principal components of the data were plotted\cite{novembre2008genes}. A diagram of the analogy between geostatistical kriging and complex trait prediction is shown in Figure \ref{kriging-flow}.

Methods of complex trait prediction using whole-genome data were pioneered by Meuwissen et al. \cite{meuwissen2001prediction} who proposed to predict the genetic effects on phenotypes as the sum of marker effects $\sum_{l=1}X_{il}\hat{\beta_l}$, where $X_{il}$ is the genotype of individual $i$ for marker $l$ and $\hat{\beta_l}$ is the estimated effect size of marker $l$.  One early application to human disease risk used a polygenic score derived from the logistic regression effect sizes of GWAS SNPs with P-values below a chosen threshold in a training set to predict a significant proportion of the risk of schizophrenia and bipolar disorder in test sets\cite{purcell2009common}. The training set was a large schizophrenia GWAS and the testing sets included multiple additional schizophrenia, as well as bipolar disorder, datasets. A similar polygenic score approach was used to predict risk of the seven diseases within the WTCCC, by dividing each sample into training sets and test sets\cite{evans2009harnessing}. When implementing polygenic score for trait prediction, the single-marker association effect sizes of SNPs that meet a chosen threshold from the training set are multiplied by the genotypes (0,1,2) of each respective SNP in the test set and the sum taken to generate a predicted phenotype value (polygenic score) for each individual. Performance of the polygenic score model is then assessed by comparing predicted to observed values. Principal components or any other known covariates that associate with the phenotype can be included in the polygenic score model to further improve prediction.

Additional whole-genome prediction (WGP) methods, which are reviewed in de los Campos et al.\cite{de2010predicting} and Abraham et al.\cite{abraham2013performance}, include penalized estimation methods such as Least Absolute Shrinkage and Selection Operator (LASSO) \cite{tibshirani1996regression}, Ridge Regression \cite{hoerl1970ridge}, and Elastic Net \cite{zou2005regularization,abraham2013performance} and the Bayesian versions of each method \cite{park2008bayesian,perez2010genomic}. These methods use different penalty functions or different types of priors to overcome the infeasibility of estimating marker effects through ordinary least squares regression in typical GWAS datasets with much larger numbers of markers than individuals. These methods prevent over-fitting and may reduce the mean-squared error of estimates and predictions\cite{de2010predicting}. As an example, Vazquez et al. used Bayesian LASSO to structure the prior density of marker effects in their Bayesian regression WGP model of skin cancer risk\cite{vazquez2012comprehensive}. Their best prediction model, which  included 41K genome-wide SNPs, had an area under the receiver operating characteristic curve (AUC) of 0.635, 18.9\% higher than that of the baseline model, which just included non-genetic covariates\cite{vazquez2012comprehensive}. 

The idea of using genetic similarity to generate prediction dates back to Fisher \cite{fisher1918} and Wright \cite{wright1921systems}. These ideas were formalized as best linear unbiased prediction (BLUP) approaches based on multivariate normal processes by Henderson, Goldberger and others\cite{goldberger1962best,henderson1950estimation,henderson1975best}. Originally, these methods used pedigree based similarity matrices but with the advent of affordable high throughput genotyping several authors have used similarity measures computed using larger numbers of genetic markers\cite{de2009reproducing,habier2007impact,janss2012inferences,lynch1999estimation,makowsky2011beyond,ober2012using,vanraden2008efficient,yang2010common}. These approaches are commonly called G-BLUP for genomic-BLUP. Unlike the polygenic score methods, G-BLUP regresses phenotypes on hundreds of thousands of markers simultaneously using a genetic relationship matrix (GRM) derived from the marker data. While most applications of G-BLUP have assumed all measured genotypes affect the phenotype with normally distributed effect sizes and thus include all markers in the GRM, a newer approach by Zhou et al. combines G-BLUP with a sparse regression model that allows for a small proportion of markers with large effect sizes and estimates the most likely model for a particular phenotype from the data\cite{zhou2013polygenic}.

The equivalence between Kriging and the BLUP methods used in the animal breeding and quantitative genetics fields has been demonstrated \cite{harville1984interpolation,robinson1991blup}. Ridge regression is equivalent to standard Kriging/BLUP, but does not have the dimension reduction advantage that the latter offers. Kriging in genomic prediction has been previously used, but it was restricted to simulation studies of genetic similarities\cite{ober2011predicting}. Based on whole genome simulations, Ober et al. reported that using Mat\'ern functions (class of special functions commonly used in geostatistics) to scale the genetic relatedness measure works better than standard measures of relatedness in the presence of dominance and epistatic effects\cite{ober2011predicting}.

The main novelty of our approach is the extension of the Kriging framework for the integration of multiple omic data. Furthermore, this framework allows easy integration of prior information on the function of the variants by partitioning the genome and giving more weight to different subsets based on functional evidence. For example, known loci can be given more weight than the rest of the genome, or subsets of the genome with regulatory evidence of affecting gene expression (eQTLs) can be given larger weight. It differs from standard Kriging/BLUP in that it is not necessarily tied to an additive genetic/genomic model. Instead of using the maximum likelihood method to estimate the parameters we use a more pragmatic approach where we maximize the cross validated prediction performance. In this sense, our approach is  closely related to the semi-parametric models using reproducing kernel Hilbert space (RKHS) regression proposed by Gianola et al.\cite{gianola2006genomic} and de los Campos et al. \cite{deloscampos2010semiparametric} for WGP. To our knowledge, our method is the first to integrate multiple omics data using these semi-parametric methods based on similarity measures. Furthermore, our intuitive connection between Kriging and BLUP should allow investigators less familiar with the quantitative genetics field to appreciate the usefulness of the approach and encourage them to adopt these methods for their specific analyses.

In sum, we propose a novel systems approach to predict complex traits, which leverages and integrates similarity in genetic, transcriptomic,  and/or other large scale omics data. Here, we describe our OmicKriging method, apply the method to several human complex traits (cellular growth, clinical statin response and seven WTCCC diseases), and provide an R package for implementation.

\section*{Results}
In implementing OmicKriging, the first step is to construct a similarity matrix or similarity matrices for use in the prediction. Such matrices may include a genetic relationship matrix (GRM) from a set of SNPs, a gene expression correlation matrix (GXM) from gene expression data, or any other correlation matrix derived from omics data. To predict the phenotype of an individual, the weighted average of the training set individuals' phenotypes is calculated. In the case of SNP data, the weights are comprised of the GRM and pairwise genotype similarity of the unknown individual with the genotypes of those with observed phenotypes. Each similarity matrix can be tested individually or in a given weighted combination for phenotype predictive performance. When using a single omic component, we tested weights between 0 and 1 (i.e. 0.1, 0.2, 0.3, ..., 1 for the omic component and 1-weight for the environmental component) to find the weight that produced optimal prediction. When two omic components (e.g. GRM and GXM) were combined, we performed a grid search to find the optimal prediction weights $w_1$ and $w_2$, such that $w_1 + w_2 \leq1$ (1-$w_1$-$w_2$ for the environmental component). The optimal weights for each omic component depend on the genetic architecture of the phenotype.  

In applying OmicKriging, we used a 16-fold cross-validation approach with individuals assigned to the 16 subsets at random and repeating the procedure 500 time to assess the sampling variability of the prediction performance (see Methods). For quantitative traits, we computed the coefficient of determination R$^2$ (equivalent to the square of the correlation) between the predicted and true values of the phenotype to assess prediction performance and for case/control traits, we computed the area under the receiver operating characteristic curve (AUC). Note that in the OmicKriging method, weighted averages are computed out-of-sample (only an individual's genotypes or gene expression levels are used to compute pairwise similarities) and should not be compared to reports where the R\textsuperscript{2} is computed using parameters estimated from the same data (in-sample).

\subsection*{Cellular phenotype applications}
To assess the predictive performance of OmicKriging,  we used the intrinsic growth rate (iGrowth) phenotype derived from multiple proliferation measurements in the commonly used HapMap lymphoblastoid cell lines (LCLs)\cite{im2012mixed}. This phenotype is relevant since it has been shown that genes associated with proliferation are strong prognostic factors in several types of cancers \cite{Starmans:2008, Damasco:2011, Dai.proliferation.breast:2005,Rosenwald.lymphoma:2003} and such genes are differentially expressed in most cancer tissues \cite{Whitfield:2006,Ross:2000, Rhodes:2004}. In addition, our unpublished data indicate that predicted drug-induced growth inhibition has predictive power on clinical response to drugs. iGrowth values from 99 LCLs from the HapMap CEU (Northern and Western European ancestry from Utah) and YRI (Yoruba from Ibadan, Nigeria) populations were used in the analysis. We tested common SNPs, gene (mRNA) expression levels, and microRNA expression levels for iGrowth predictive ability. The GRM was generated from 2.7 million common SNPs (minor allele frequency $>$ 0.05) from HapMap\cite{frazer2007second} using GCTA\cite{yang2011gcta} embedded in our OmicKriging R package. The GXM was generated from 13,080 transcript clusters from a previous genome-wide gene expression analysis\cite{zhang2008evaluation} as simply the correlation matrix of the expression data as described in the Methods. We also obtained expression measurements for 201 microRNAs\cite{gamazon2012genetic} and generated a microRNA expression similarity matrix (MXM). 

\begin{figure}[!ht]
\begin{center}
\includegraphics[width=5in]{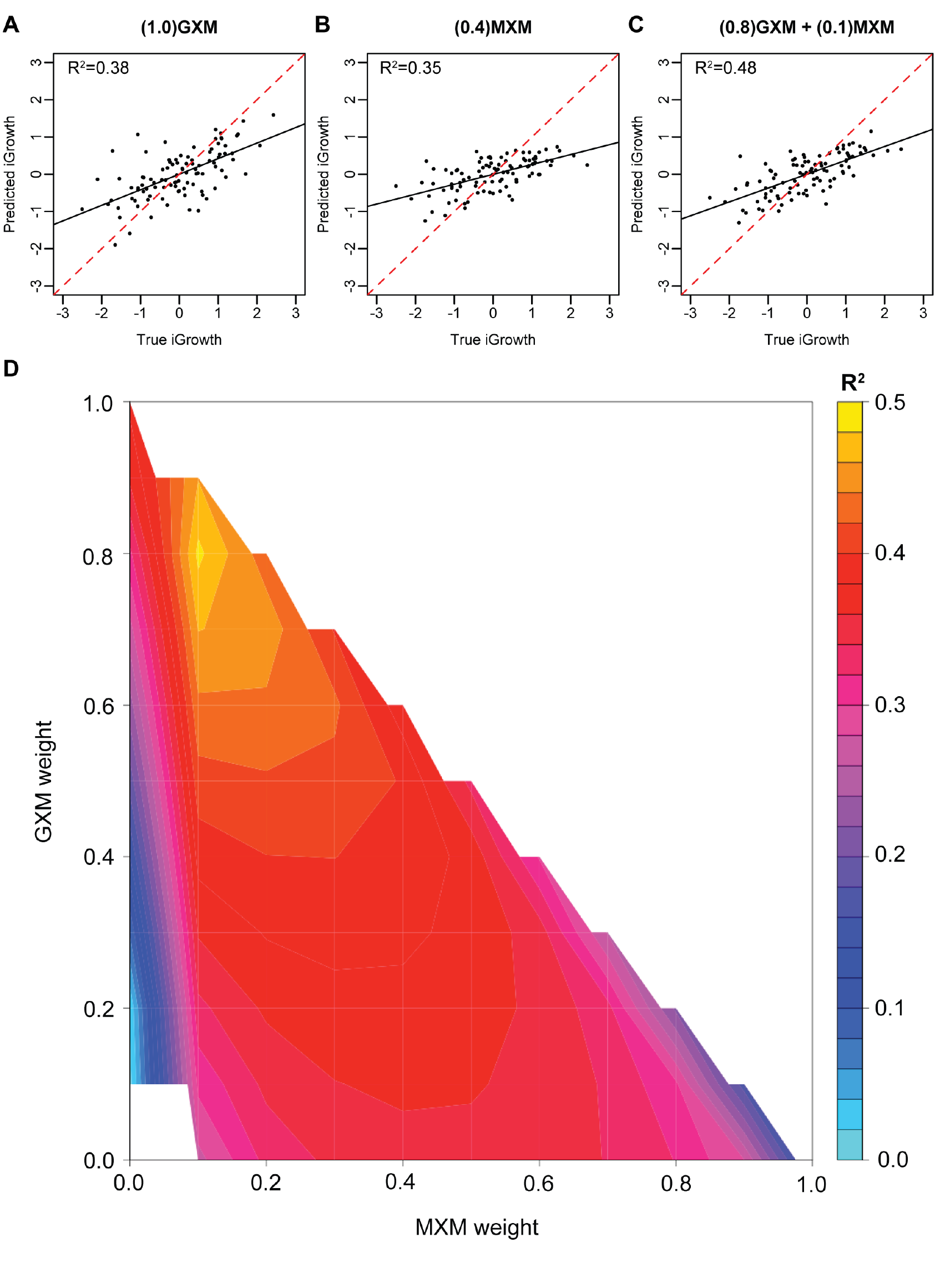} 
\end{center}
\caption{
{\bf iGrowth prediction using OmicKriging.} Predicted versus true iGrowth (n=99) using (A) the optimally weighted gene expression matrix (GXM) alone, (B) the optimally weighted microRNA expression matrix (MXM) alone, and (C) the optimally weighted combination of the two matrices from the grid search. The solid black lines represent the slopes of the regression between the predicted and true values. The red dashed lines are the identity lines representing perfect prediction (slope 1, intercept 0). (D) Results of the grid search which shows that the best iGrowth prediction correlation (R$^2=0.48$ [0.45, 0.52]) was obtained with (MXM, GXM) matrix weights of (0.1, 0.8). The R$^2$ values presented in the contour plot are the mean values from 500 random samplings of the data into 16 cross-validation folds.   }
\label{3matrices}
\end{figure}

We first tested each single similarity matrix for predictive ability using OmicKriging. At every weight $w_\text{GRM}$ attempted ($w_\text{GRM}=0.1, 0.2, ..., 1$ and $w_\epsilon=1-w_\text{GRM}$), the GRM did not show any predictive power (i.e. the correlation between the predicted and true iGrowth values did not differ from zero). However, for the GXM alone, the optimal prediction correlation was R$^2=0.38$ [0.34, 0.43] when $w_\text{GXM}=1$ (Figure\ref{3matrices}A). The 95\% confidence interval of each prediction [in brackets] was determined by 500 permutations of randomly partitioning the data into training and test sets as described in the Methods.  In addition, for the MXM alone, the optimal prediction correlation was R$^2=0.35$ [0.32, 0.38] when $w_\text{MXM}=0.4$ and $w_\epsilon=0.6$ (Figure\ref{3matrices}B). We then performed a grid search to determine if combining the GXM and MXM similarity matrices improved the iGrowth prediction. The optimal prediction from the grid search increased the correlation to R$^2=0.48$ [0.45, 0.52], when $w_\text{GXM}=0.8$, $w_\text{MXM}=0.1$, and $w_\epsilon=0.1$ (Figure\ref{3matrices}C-D). The non-overlapping confidence intervals indicate that combining genome-wide expression data improved the iGrowth predictive power of OmicKriging over using either the GXM or MXM alone.  

We developed a baseline model similar to a polygenic score model, but with the top gene and microRNA  expression associations rather than SNP associations, to compare to our OmicKriging results. Specifically, using the entire sample, we determined that 255 genes and 14 microRNAs associated with iGrowth by univariate linear regression after Bonferroni correction for multiple tests. Then, we performed 16-fold cross-validation by determining the top 255 genes and top 14 microRNAs by univariate linear regression in each training set and using the effect sizes to predict iGrowth in each test set. In addition to the top 255 genes and top 14 microRNAs from each training set, we also included the first 10 principal components derived from the genotype data in each multivariate prediction model. We repeated the cross-validation procedure 500 times to generate a confidence interval. The baseline model was unable to predict iGrowth, R$^2=0.0038$ [-0.010, 0.064]. Therefore the maximum R$^2=0.48$ obtained by OmicKriging represents a vast improvement in iGrowth prediction over the baseline model of top expression associations (Table \ref{tab:igrowth}).

\subsection*{Clinical phenotype applications}

\paragraph{Clinical statin response}
We also applied OmicKriging to the Cholesterol and Pharmacogenetics (CAP) Simvastatin Study, which contains both genome-wide genotype and expression data\cite{barber2010genome,medina2012rhoa,simon2006phenotypic}.  DNA samples were either genotyped on the Illumina HumanHap 300K beadchip or the Illumina HumanHap 610K-Quad beadchip. A total of 562 individuals had their change in low-density lipoprotein cholesterol (dLDLC) after simvastatin treatment measured and also passed genotyping quality control. Additional SNP genotypes were imputed into these samples using data from the 1000 Genomes Project\cite{autosomes2012integrated}. To assess potential predictive ability of the SNPs, we estimated the heritability of dLDLC captured by the 8.7M imputed SNPs to be 0.69 (SE 0.61) using GCTA\cite{yang2011gcta}. Because of the large standard error obtained when all the SNPs were used, we sought to obtain a significant estimate of heritability by using a subset of SNPs known to be involved in statin response from a separate study. The Heart Protection Study identified 45 SNPs (Tables 2 and S3 in Hopewell et al.) to be involved in LDLC response to simvastatin\cite{hopewell2013impact}. We used SNPs within 50kb of these 45, for a total of 10,925 SNPs, and obtained a heritability estimate for dLDLC of 0.16 (SE 0.079, $P=0.01$). The estimated heritability is quite low but the fact that it is significantly different from zero indicated that significant predictive power might be attainable. When we applied OmicKriging to the CAP SNP data for dLDLC prediction, the optimal prediction correlation was  R$^2=0.037$ [0.026, 0.049] when $w_\text{GRM}=0.8$ using a single GRM calculated from the Hopewell 50kb SNPs. Adding a second GRM of all imputed SNPs and performing a grid search did not improve the dLDLC prediction.

To compare our OmicKriging method to existing WGP methods, we used polygenic score and attempted several sets of top SNPs for prediction of dLDLC using a 16-fold cross-validation approach. The best prediction had R$^2=0.016$ [0.0048, 0.030] when the top 1000 SNPs from each training set were used to predict the respective test set, which is less than half the prediction R$^2$ obtained with OmicKriging (R$^2=0.037$, Table \ref{tab:dldlc}).  

Of the 562 individuals in the CAP study, 461 of them also had gene expression measurements from LCLs derived from their blood. Patient LCLs were exposed to simvastatin or vehicle control (baseline) for 24H and then gene expression was measured using the Illumina Ref8v3 beadchip\cite{medina2012rhoa}. The differences between the treated and baseline expression values for each individual were used to generate a GXM. Including all 12,951 expressed genes in the GXM and performing a grid search with the Hopewell GRM did not improve the dLDLC prediction over the GRM alone. Like we did for the SNPs, we then chose to only include genes from candidate pathways known to be involved in lipid metabolism and inflammation in the GXM in an attempt to improve prediction. Combining a GXM of the 28 expressed genes from the PID RHOA REG PATHWAY from the Molecular Signatures Database (MSigDB)\cite{liberzon2011molecular} with a GRM of the Hopewell SNPs produced an optimal R$^2=0.025$ [0.014, 0.036] when $w_\text{GXM}=0.3$ and $w_\text{GRM}=0.6$. This prediction was not significantly different from using the Hopewell GRM alone with this reduced sample size (R$^2= 0.024$ [0.015, 0.035], when $w_\text{GRM}=0.9$). We also developed a baseline model using the 45 SNPs reported by Hopewell et al. combined with the first 10 principal components and the top 396 gene expression levels in a polygenic score prediction, but the R$^2$ was negligible (R$^2=-0.00014$ [$-$0.0022, 0.0070]). Thus, combining the genotype and gene expression data did not improve the prediction (Table \ref{tab:dldlc}).

\paragraph{WTCCC diseases}

\begin{figure}[!ht]
\begin{center}
\includegraphics[width=6in]{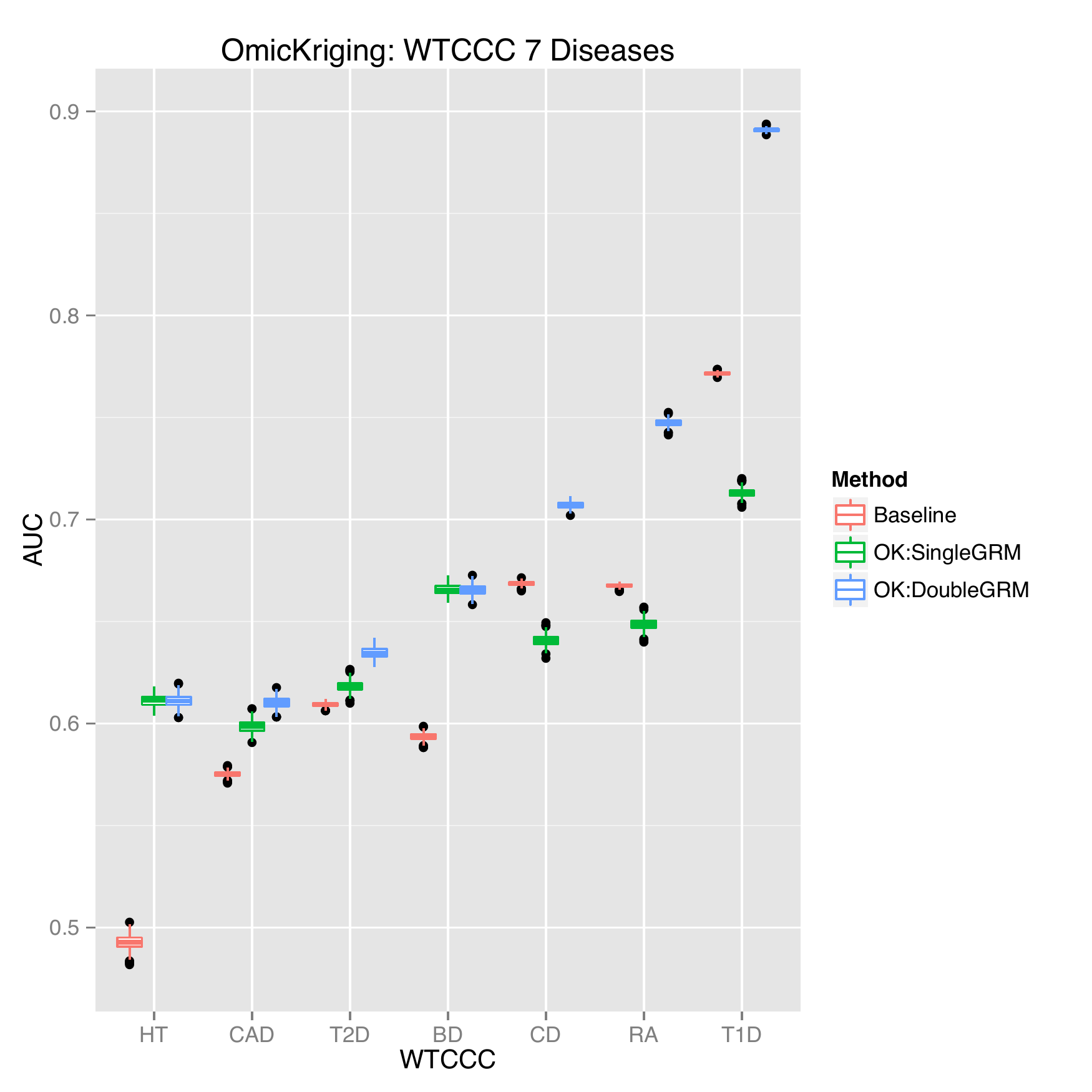} 
\end{center}
\caption{
{\bf OmicKriging prediction performance for WTCCC disease risk prediction.} Boxplots of the area under the ROC curve (AUC) from 500 permutations of the data for two implementations of OmicKriging for each disease from the WTCCC: a single common SNP genetic relationship matrix (OK:SingleGRM) and two optimally weighted GRMs  of common SNPs and known loci (OK:DoubleGRM) for the predictions. The double GRM AUCs dramatically increase for CD, RA, and T1D, which are known to have multiple associated loci of relatively large effect. The known loci were obtained from studies that did not include the WTCCC data to avoid over-fitting. For comparison, we also show AUC results from 500 permutations of the polygenic score method using genome-wide significant loci with 10 principal components (Baseline). Both OmicKriging implementations outperform the baseline model for HT, CAD, T2D, and BD. The OK:DoubleGRM greatly outperforms the baseline model for CD, RA, and T1D.  BD=bipolar disorder, CAD=coronary artery disease, CD=Crohn's disease, HT=hypertension, RA=rheumatoid arthritis, T1D=type 1 diabetes, T2D=type 2 diabetes.    }
\label{auc}
\end{figure}

To test the predictive power of OmicKriging using larger clinical datasets, we turned to the seven diseases of the Wellcome Trust Case Control Consortium (WTCCC)\cite{burton2007genome}. First, for each case/control dataset, all genotyped common SNPs (approximately 400,000) were used to generate a GRM. OmicKriging was run using these single GRMs (equivalent to G-BLUP) and case/control prediction was successful (AUC $>$ 0.50 which would be considered random guess) for all seven diseases as has been shown previously\cite{zhou2013polygenic}. In our analysis, mean areas under the ROC curve (AUCs) ranged from 0.598 [0.593, 0.604] when $w_\text{GRM}=0.4$ for coronary artery disease to 0.713 [0.709, 0.717] when $w_\text{GRM}=0.4$ for type 1 diabetes (Figure \ref{auc}, Table \ref{tab:disease}). The 95\% confidence intervals [in brackets] were determined by randomly partitioning the data 500 times into 16 subsets  and performing cross-validated prediction on every random partition to generate a distribution of 500 AUC values. While we did perform a grid search to determine the best AUC for each WTCCC disease, optimization typically resulted in minimal improvement. That is, the optimized $w_\text{GRM}$ did not improve the AUC greater than 0.02 over the default $w_\text{GRM}=1$.

In an attempt to further improve the prediction by integrating existing information on variants from previous studies, we generated a second GRM for each disease using the SNPs within 100kb of known loci (identified outside of WTCCC studies for each disease) listed in the The National Human Genome Research Institute GWAS catalog\cite{hindorff2009potential} and the Database of Genotypes and Phenotypes (dbGAP)\cite{mailman2007ncbi}. The optimal double GRM improved the predictive power of OmicKriging over using just the single common-SNP GRM alone slightly for coronary artery disease and type 2 diabetes and dramatically for Crohn's disease, rheumatoid arthritis, and type 1 diabetes (Figure \ref{auc}). The type 1 diabetes prediction showed the largest improvement when the second GRM was added: the AUC increased from 0.713 to 0.891 [0.889, 0.892] (Figure \ref{auc}, Table \ref{tab:disease}).

We compared OmicKriging to a baseline model that uses only genome-wide significant SNPs and the first ten principal components to calculate predicted phenotypes by the polygenic score method. Both OmicKriging models (single and double GRM) outperformed the baseline model for coronary artery disease, hypertension, type 2 diabetes, and bipolar disorder (Figure \ref{auc}, Table \ref{tab:disease}). The OmicKriging double GRM model greatly outperforms the baseline model for Crohn's disease, rheumatoid arthritis, and type 1 diabetes. The greatest differences between models were seen for rheumatoid arthritis and type 1 diabetes, where the OmicKriging mean AUCs were 0.08 and 0.12 higher than the baseline polygenic score mean AUCs, respectively (Figure \ref{auc}, Table \ref{tab:disease}).

\section*{Discussion}

We propose a novel systems approach to predict complex traits, which leverages and integrates similarity in genetic, transcriptomic  and/or other large scale omics data. We translate the genomic similarity into phenotypic similarity using a method called Kriging, commonly used in geostatistics and machine learning. Here, we construct the genomic similarity using a linear combination of omic similarity matrices (in addition to the environmental component), but more general similarity matrices can be used. Given the similarity matrix, the prediction is obtained by simply computing a weighted average of the phenotype of individuals in the training set. The weights are provided by the Kriging method, which can be loosely interpreted as a converse of the linear regression method. Our method is a fast, simple and flexible approach to polygenic, and more generally poly-omic, prediction.
 
The key component of the method is the choice of the similarity matrix. If we assume certain conditions such as additive poly-omic models (as described in the Methods section) it is possible to estimate the weights for each omic component as well as the environmental component using a (restricted) maximum likelihood approach or Bayesian generalizations. However, some of the modeling assumptions are quite stringent and we opted for a more pragmatic approach in which we use the weights that maximize predictive performance. In this manuscript, we use a grid search approach to find reasonably close to optimal weights for each similarity matrix component (GRM of known large effect SNPs, GRM of all SNPs, GXM, etc.).  

Using mRNA (GXM) and microRNA (MXM) data in HapMap LCLs, we predict the cellular intrinsic growth phenotype with an out of sample R$^2$ of 0.48 with just 99 samples. For comparison, previous studies using similar datasets had reported R$^2$ values of the same magnitude but were computed in-sample, which is known to be highly upward-biased. The combined mRNA-microRNA prediction R$^2$ was 0.10-0.13 higher than using either the GXM or MXM alone. Given that 30\% of gene expression levels associate with the intrinsic growth phenotype (FDR$<$0.1)\cite{im2012mixed}, it is perhaps not surprising that gene expression correlations are most useful in predicting this phenotype. However, using just the most significantly associated expression levels as was done in the baseline polyscore model, was not predictive. These results stress the importance of considering other potential biomarkers besides genotype, because the GRM was not helpful in predicting this particular phenotype at this sample size. LCLs from populations of both African and European descent were used in this analysis. 

The issue of population stratification has slightly different implications in the prediction context than it does in the GWAS estimation context. If our goal is to achieve the best prediction possible, we are not concerned whether the predictive power comes from the genomic component or environmental and populations stratification components that are confounded with the genomic component. Naturally, we have to be aware of this limitation and not to expect similar levels of accuracy when predicting in different population mixtures.

Because the dataset contained genotype and gene expression data, we used the CAP simvastatin study to test the potential predictive ability of OmicKriging applied to a clinical phenotype with multiple types of omics-level data. While the out-of-sample prediction correlations were not large with OmicKriging, likely due to limited sample size, the best SNP-based prediction R$^2$  was more than twice as high as that obtained from the polygenic score method (R$^2=0.037$ vs. R$^2=0.016$). While inclusion of expression data did not improve prediction over genetic variants alone, the CAP study examples demonstrate how OmicKriging can be applied to multiple types and subsets of omics data for phenotypes with such data available.

We also successfully predicted seven clinical disease risk phenotypes with OmicKriging. For WTCCC disease prediction, we show that our OmicKriging method yields performance similar to or better than polygenic score. In addition, OmicKriging when restricted to genotypic data performs just as well as the computationally more intensive BSLMM method\cite{zhou2013polygenic} (Table \ref{tab:disease}). The average OmicKriging double GRM run time on a Xeon E5345 processor is 14 minutes, whereas the average GEMMA software\cite{zhou2012genome} run time for BSLMM is 28 hours on the Xeon L5420 processor, which is a slighly newer, but comparable processor\cite{zhou2013polygenic}. Most of the BSLMM run time is used for the Markov chain Monte Carlo (MCMC) iterations, whereas in OmicKriging, we specify the sparse effects (known GWAS loci) before the run. For the known autoimmune diseases (Crohn's disease, rheumatoid arthritis, type 1 diabetes), adding a second GRM of known loci increased the prediction AUC values over a single common SNP GRM alone to AUC values slightly higher (not significant) than those obtained by BSLMM (Table \ref{tab:disease}).\cite{zhou2013polygenic}. These autoimmune diseases are known to have multiple associated loci of relatively strong effects, so this prior knowledge was used to improve prediction performance\cite{visscher2012five}. Unless replicated in an independent study, the results from the WTCCC data were not used to select top SNPs to avoid overfitting the data. The potential for prediction improvement by differentially weighting markers based on trait specific information was explored in simulated data by Zhang et al. using BLUP models\cite{zhang2010best}.

While we recognize that assuming that a binary trait is continuous is not statistically optimal, we do so here in our initial modeling for computational reasons, as have others\cite{lee2011estimating,zhou2013polygenic}. Linear probability models for binary outcomes are considered to be adequate approximations when the proportion of cases (and controls) exceed 25\% given the approximate linearity of the logit or probit functions near the origin \cite{zhou2013polygenic}. It has been reported that the gain in statistical efficiency is hard to realize because of the added computational burden and consequent loss in numerical accuracy \cite{zhou2013polygenic,visscher1996mapping}. Unlike us, Vazquez et al. used a probit link model for skin cancer incidence, but needed to restrict their analysis to only 41K SNPs due to computer memory limitations\cite{vazquez2012comprehensive}. In their dataset, the linear probability model would have been less appropriate since the proportion of cases was between 11-24\%.

While the equivalence between Kriging and BLUP methods used in the quantitative genetics field has been demonstrated \cite{harville1984interpolation,robinson1991blup}, our OmicKriging method differs from standard Kriging/BLUP in that we do not prespecify a model. We use an additive poly-omic model to motivate the main functional form of the similarity matrix but find the weights for each omic component (and the environmental component) by maximizing the cross-validated prediction performance. Thus our approach is  closely related to the semi-parametric models using RKHS regression proposed by Gianola et al.\cite{gianola2006genomic} and the kernel averaging approach proposed by de los Campos et al.\cite{deloscampos2010semiparametric} for WGP. We use an additive poly-omic model to motivate the structure of the composite similarity matrix, but have ignored the cross-correlation terms between different omic components by assuming independence of the effects from different sources. This is a very restrictive assumption, but including the cross-correlation term is a complex topic that merits further investigation. It may be possible to use eQTL information to restrict the number of non-zero correlations, but even then further assumptions must be made to be able to estimate the needed parameters given available data and computational limitations.

The main novelty we add to these Kriging/RKHS methods based on similarity measures is the integration of multiple types of large-scale omic data. Our OmicKriging framework allows easy integration of prior information on the function of variants or pathways from heterogenous sources by giving more weight to different subsets of SNPs or gene expression levels without the sometimes heavy computational burden of Bayesian approaches. For example, eQTLs could be given more weight than SNPs with no evidence of regulatory function. In addition, expression levels in candidate pathways related to a particular phenotype could be given more weight and tested for predictive performance. Furthermore, our analogy comparing Kriging for rainfall to OmicKriging for phenotypic prediction and our applications to omic data provide a more intuitive understanding of the method than a formal mathematical definition. This should allow a broader audience to appreciate the usefulness of OmicKriging and to adopt the approach. We provide an R package called OmicKriging and a tutorial explaining how to implement the kriging algorithm at \url{http://www.scandb.org/newinterface/tools/OmicKriging.html}.













\section*{Methods}

\subsubsection*{OmicKriging Approach}

We propose to use an extension of the Kriging framework to integrate different omic data as well as prior information on function such as existing GWAS studies, eQTL information (genetic markers associated with gene expression levels), regulatory evidence such as provided by ENCODE studies, etc. We build the similarity matrix as a linear combination of the similarity matrices from each omic component where the coefficients or weights for each component are chosen so that prediction performance is maximized. Given a similarity matrix, the usual Kriging formulas are used to compute the predicted values. Prediction is performed by randomly partitioning the samples into 16 subsets and using each subset as the testing set and the remaining 15 sets as the training set. This is repeated 500 times to assess the sampling variability. The correlation squared between the true and predicted values are used as performance measures. For binary/disease traits we use the area under the receiving operating  curve. In this work, we assume a linear probability model for the disease status following Lee et al. \cite{lee2011estimating} and Zhou et al. \cite{zhou2013polygenic}.

\paragraph{Omic Similarity Matrix}

For each omic dataset used for prediction we compute the corresponding similarity matrix. For convenience, we will denote the similarity matrix constructed from genetic data as GRM, the one constructed by mRNA expression profile data as GXM, and the ones constructed with microRNA expression data as MXM. We assume that the environmental component is independent across individuals and is represented by the identity matrix, $\mathbb{I}$.
In the current implementation of the OmicKriging R package, we are computing the similarity matrix for genetic data by invoking the GCTA \cite{yang2011gcta} software, whereas for other omic data we compute the similarity matrix directly in R. More specifically, the $ij$ component of the GRM is computed as 
\[ \frac{1}{M}\sum_{l=1}^M \frac{(X^G_{il} - 2p_l)(X^G_{jl} - 2p_l)}{2 p_l (1-p_l)}\] 
and the $ij$ component of the other omics data is computed as 
\[\sum_{l=1}^L \frac{(X^O_{il} - \bar{X}^O_i)(X^O_{jl} - \bar{X}^O_j)}{ \sqrt{\sum_k(X^O_{ik} - \bar{X}^O_i)^2 \sum_k{(X^O_{jk} - \bar{X}^O_j)^2} } }\] 
where $i$ and $j$ denote individuals, $X^G_{il}$ is the number of reference alleles of individual $i$ at marker $l$, $p_l$ is the reference allele frequency of marker $l$, $X^O_{il}$ is the level of omic marker $l$ ($l$ is a dummy index and there is no one to one correspondence between genetic and omic markers indices), $M$ is the number of genetic markers,  $L$ is the number of genes or omic markers, $\bar{X}^O_i = \sum_k X^O_{ik}/L$, and $ \bar{X}^O_j = \sum_k X^O_{jk}/L$. 

By using the correlation without prior centering and standardizing the gene expression and other continous omic traits, we are effectively giving more weight to the traits that have larger variance. The effects of different choices of the similarity matrix on the prediction accuracy will be investigated in future work.

\subsubsection*{Optimal similarity matrix under an additive model}

A key component of the success of OmicKriging is understanding which proximity measures translate best into phenotypic similarity. The optimal similarity matrix depends on the underlying genetic and epigenetic architecture of the complex trait.

We will use an additive poly-omic model to motivate our choice of the similarity matrix. The phenotype for individual $i$, $Y_i$ (a scalar) is represented as
\[Y_i = a + G_i + T_i + O_i + \cdots + \epsilon_i \]
where
\begin{itemize}
	\item $a$ is a constant,
	\item $G_i = \sum_{l=1}^{M}\beta_{l}^G X_{il}^{G}$ is the additive genetic component (assumed to be known), $\beta_l^G$ is the effect size of the standardized genotype $X_{il}^G$, and $M$ is the total number of genetic markers,
	\item $T_i = \sum_{l=1}^{L}\beta_{l}^T X_{il}^{T}$ (assumed to be known),  $\beta_l^T$ is the effect size of the standardized gene expression $X_{il}^T$ and $L$ is the total number of genes, 
	\item $O_i = \sum_{l=1}^{L'}\beta_{l}^O X_{il}^{O}$ is the additive (other) omic component (assumed to be known), $\beta_l^O$ is the effect size of the standardized omic level $X_{il}^O$, and $L'$ is the total number of omic markers,
	\item and $\epsilon_i$ is a noise term (iid, independent and identically distributed).
\end{itemize}
For notational convenience let us define $X_i$ without a superscript to denote all three omic data such that $X_{il} = X^G_{il}$ if $l \le M$, $X_{il} = X^T_{i,l-M}$ if $M<l\le M+L$ and $X_{il} = X^O_{i,l-M-L}$ if $M+L<l\le M+L+L'$ and similarly for coefficients $\beta$'s such that
$\beta_{l} = \beta^G_{l}$ if $l \le M$, $\beta_{l} = \beta^T_{l-M}$ if $M<l\le M+L$ and $\beta_{l} = \beta^O_{l-M-L}$ if $M+L<l\le M+L+L'$.

We assume a random effects model for the $\beta$'s. For convenience we also assume that the $X$'s have been centered and standardized. 
If we further assume that the betas are independent, i.e. that cov($\bm\beta$) = $\sigma_\beta^2 \mathbb{I}$ then the covariance matrix of the n-vector $Y$ will have components
\begin{eqnarray}
 \Sigma_{i,j} &=& \theta_G ~ \sum_{l=1}^{M} X^G_{il}X^G_{jl} + \theta_T ~  \sum_{l=1}^L X^T_{il}X^T_{jl} + \theta_O   \sum_{l=1}^{L'} X^O_{il}X^O_{jl} + \theta_\epsilon\delta_{ij}
\end{eqnarray}
where $\delta_{ij}$ is the kronecker delta (1 if $i=j$ and 0 otherwise) and $\theta_G$, $\theta_T$, and $\theta_O$ are non negative.
If all modeling assumptions were met and we assumed normality of the environmental term, this covariance matrix should be used as the similarity matrix to compute the best linear unbiased prediction (BLUP). However, these assumptions are quite strong and do not account for correlations of between marker effects,  gene-gene interactions, gene-environment interactions, etc. Thus, we adopt a pragmatic approach in which we use a combination of the covariance matrices for each omic component but allow the weights $\theta$ to vary and pick the combination that provides the best predictive performance.

Independence assumption for all betas is clearly too restrictive. The effect size of genetic marker $X_{il}^G$, $\beta_l^{G}$, that influences gene expression level $X_{ik}^T$ is likely to be correlated with $\beta_k^T$. For unconstrained values of the cov($\beta_k$,$\beta_l$) the covariance matrix has the form
\begin{eqnarray*}
 \Sigma_{i,j} &=& \theta_G ~ \sum_{l=1}^{M} X^G_{ik}X^G_{jk} + \theta_T ~  \sum_{l=1}^L X^T_{ik}X^T_{jk} + \theta_O   \sum_{k=1}^{L'} X^O_{ik}X^O_{jk} + \theta_\epsilon\delta_{ij} \\
 & + &  ~  \sum_{k \ne l} \text{cov}(\beta_k,\beta_l) X_{ik}X_{jl} 
\end{eqnarray*}
In case prior expression quantitative trait loci (eQTL, genetic markers that have an effect on gene expression traits) information is available, it may be possible to restrict the non zero cross-correlation terms to known eQTL pairs ($X^G_{l}$,$X^T_k$). Additional restrictions in the values of the $\text{cov}(\beta_l,\beta_k)$ must be imposed to be able to characterize them given existing data and care must be taken to preserve positive definiteness of $\bm{\Sigma}$ (all eigenvalues must be $> 0$). This is a complex topic that merits further research. In this paper, to keep computations within reach, we ignore the cross-correlation terms and find the coefficient thetas that maximize prediction performance.

\paragraph{Composite Similarity Matrix}

Based on the form of the optimal similarity matrix under an additive poly-omic model, we propose to use a composite similarity matrix that integrates different omic components to be used for Kriging that is a linear combination of each component similarity matrix ($S_s$) 
\[\Sigma = \theta_1 S_1 + \theta_2 S_2 + \theta_3 S_3 + \cdots + (1-\theta_1-\theta_2-\theta_3 \cdots) \mathbb{I}\]
where the weights $\theta$'s will be determined as the ones providing optimal prediction. All coefficients $\theta$ are constrained to be non-negative. The environmental component is known as the nugget term in geostatistical applications.

\subsubsection*{Kriging Formula}

Within the kriging framework, the predicted phenotype of a test individual is computed as the weighted average of the phenotype of the individuals in the training set.
\begin{equation}
 \text{Prediction}(Y_{\text{new}}) = \omega_1 Y_1 + \omega_2 Y_2 + \cdots + \omega_n Y_n
\end{equation} 
where the weights $\omega_i$ are a function of all $n (n+1)/2$ pairs of similarity measures. In the simplest case where no covariates are needed, the weights prescribed by the Kriging method are given by
\begin{equation}
	{\bm \omega} =  \bm \Sigma^{-1} {\bm \rho}
\end{equation}
where
$\boldsymbol\rho$ is the similarity vector between the test individual and the training individuals and 
$\boldsymbol\Sigma$ is the similarity matrix of the individuals in the training set\cite{cressie1993statistics}.
Covariates are easily included in the method by using the so called universal kriging approach\cite{cressie1993statistics}. Assuming there are $p$ covariates (if only the intercept is considered, $p=1$), let $\bm{z}$ be the $p$ by $1$ vector with covariates $1$ to $p$ 
corresponding to the test individual and $\mathbb{Z}$ be the $n$ by $p$ matrix with the $p$ covariates for the $n$ individuals in the training set. The weights become
\begin{equation}
	{\boldsymbol \omega} = \Sigma^{-1} \left( \bm{\rho} + \mathbb{Z} \bm{m} \right ) 
\end{equation} where
$  \bm{m} =  ( \mathbb{Z}' \bm{\Sigma}^{-1} \mathbb{Z} )^{-1}( \bm{z} - \mathbb{Z}' \bm{\Sigma}^{-1} \bm{\rho} ).$


\subsubsection*{Prediction Performance} We measure prediction performance for binary traits with the area under the receiving operator characteristic curve (AUC). For quantitative traits, we use $R^2$, the correlation coefficient between true and predicted values squared.

\subsubsection*{Grid Search}

When using a single similarity matrix, we compared the prediction performance measures for similarity matrix weights, $w_1 = 0, 0.1, 0.2, ... 1$, and environmental component weights, $w_\epsilon = 1 - w_1$. For two similarity matrices (e.g. GRM and GXM) we allow matrix weights $w_1 = 0, 0.1, 0.2, ... 1$ and $w_2 = 0, 0.1, 0.2, ... 1$, with the constraint that $w_1+w_2 \le 1$. When two similarity matrices are used, the environmental component weights are $w_\epsilon = 1 - w_1 - w_2$.

\subsubsection*{Sampling Variability}

To assess the uncertainty of the R$^2$ or AUC estimates due to sampling variability, we randomly partitioned each dataset 500 times into 16 subsets  and performed cross-validation on every random partition. That is, within each cross-validation fold, 1/16 of the data was used as the test set and 15/16 of the data was used as the training set.  This random sampling and cross-validation generated a distribution of 500 R$^2$ or AUC values for each prediction method and trait from which 95\% confidence intervals were calculated (as the 0.025 and 0.975 percentiles of the distribution of R$^2$ or AUC values). All analyses were performed using the R statistical language and environment\cite{team2005r}.

\subsubsection*{Clinical statin response analysis}

\paragraph{Genotype imputation}
Self-reporting Caucasian individuals from the Cholesterol and Pharmacogenetics Study\cite{simon2006phenotypic} were genotyped on the Illumina HumanHap 300K beadchip (n=305) or the Illumina HumanHap 610K-Quad beadchip (n=282). Prior to imputation, we performed standard GWAS quality control, removing poorly called SNPs and SNPs in Hardy-Weinberg disequilibrium (HWD, P $<$ 0.001). We also removed related individuals and outliers for heterozygosity or principal components. This left 562 individuals, who also had baseline and post-simvastatin treatment LDLC measurements, for imputation. Prior to imputation we pre-phased the genotype data using SHAPEIT\cite{delaneau2011linear} using the recommended settings. Then we used IMPUTE2\cite{howie2009flexible} to impute genotypes from the 1000 Genomes Project\cite{autosomes2012integrated} using the default settings for pre-phased data. A total of 8.7M SNPs with IMPUTE2-info scores $>$ 0.3 and minor allele frequency (MAF) $>$ 0.001 and genotypes with probabilites $>$ 0.9 were used in the heritability estimation and prediction analyses. 
\paragraph{Phenotype}
The change in low-density lipoprotein cholesterol (dLDLC) phenotype was calculated by subtracting the log-transformed mean (over the two baseline measurements) of the baseline LDLC plasma levels from the log-transformed mean (over the two visits post-treatment) of the LDLC plasma levels collected while patients were on simvastatin\cite{simon2006phenotypic}.
\paragraph{Expression pathway analysis}
To test whether the gene expression of pathways potentially related to simvastatin-induced LDLC response could predict dLDLC, we chose a few canonical pathways from the MSigDB\cite{liberzon2011molecular} to test in our OmicKriging model as proof-of-concept for future more comprehensive pathway analyses. The pathways tested for prediction ability include BIOCARTA INFLAM PATHWAY, PID RHOA PATHWAY, PID RHOA REG PATHWAY, and REACTOME CHOLESTEROL BIOSYNTHESIS. While RHOA has been previously implicated in lipid metabolism and thus makes a plausible candidate pathway, we focus on it here over many other potential lipid metabolism pathways, because recent functional work in the CAP LCLs have revealed specific effects of RHOA (ras homolog family member A) in modulating the cholesterol-lowering effects of statin\cite{medina2012rhoa}.

\subsubsection*{WTCCC disease analysis}

We performed standard quality control for all WTCCC data sets. All WTCCC data sets were merged into a single bed file where we removed all individuals recommended by the WTCCC. We removed SNPs in HWD P $<$ 0.0005 and MAF $<$ 0.01. This resulted in approximately 388K SNPs after pruning. In addition, we computed the GRM for all WTCCC and identified a pair of cases with unusually high relatedness that were not included in the WTCCC removal list. The duplicate individual was removed.  

\paragraph{OmicKriging models} We selected SNPs from dbGAP and NHGRI to be used in the double GRM model\cite{hindorff2009potential,mailman2007ncbi}. We pruned all SNPs from studies that contained WTCCC datasets. We fit OmicKriging models with single (all SNPs) and double GRMs (all SNPs and known GWAS SNPs) in 16-fold cross-validation. This fold cross-validation number was chosen because OmicKriging can be multithreaded. For these analyses, we used dual Xeon E5620 processors with 16 logical cores to run OmicKriging.

We performed a grid search (as described previously) to identify the optimal weights for single and double GRM models (Table \ref{tab:disease}). Prediction performance of all OmicKriging and baseline models was measured by area under the receiver operating characteristic (ROC) curve (AUC) with the package ROCR in R \cite{sing2005rocr}. We used the R package ggplot2 \cite{wickham2009ggplot2} to generate Figure \ref{auc}.

\paragraph{Polygenic score models}
We applied the polygenic score method by fitting the first 10 principal components and p genome-wide-significant loci jointly in 16-fold cross-validation (baseline model). Specifically, we fit $\text{Y} {\raise.17ex\hbox{$\scriptstyle\sim$}} \text{PC1} + \cdots + \text{PC10} + \text{SNP1} + \cdots + \text{SNPp}$ in each training set (15/16th of the dataset). With the remaining test set (1/16th of the dataset), the $m=$p$+10$ estimated regression coefficients ($\hat{\beta}$) are multiplied by an $n$x$m$ matrix of $n$ individuals and $m$ principal components/genotype dosages ($Z_{il}$) and each individual's predicted phenotype (polygenic score) is the sum of the respective individual's products:  $\sum_{l=1}^mZ_{il}\hat{\beta_l}$.

\section*{Acknowledgments}

We thank Peter McCullagh for useful comments on the manuscript.
The Cholesterol and Pharmacogenetics Study data used for the analyses described in this manuscript were obtained through an agreement with the investigators and the Pharmacogenomics Research Network Statistical Analysis Resource Workshop and from the database of Genotype and Phenotype (dbGaP) found at http://www.ncbi.nlm.nih.gov/gap through dbGaP accession number phs000481.v1.p1. 
We acknowledge the PARC investigators and research team, supported by NHLBI, for collection of data from the Cholesterol and Pharmacogenetics clinical trial. 
This study makes use of data generated by the Wellcome Trust Case-Control Consortium. A full list of the investigators who contributed to the generation of the data is available from www.wtccc.org.uk. 

\bibliography{kriging,kriging-more}

\section*{Figure Legends}

\newpage
\section*{Tables}

\begin{table}[!ht]
\caption{
\bf{iGrowth prediction is maximized by integration of expression data through OmicKriging.}}
\begin{tabular}{|p{1.5cm}|p{2.2cm}|p{2.2cm}|p{2.2cm}|p{2.2cm}|p{2.2cm}|}
\hline
Model: & OmicKriging GRM & OmicKriging GXM & OmicKriging MXM & OmicKriging GXM+MXM & baseline\\
\hline
mean R$^2$ [95\% CI] & 0.020 [-0.0083, 0.077] & 0.38 [0.34, 0.43] & 0.35 [0.32, 0.38] & 0.48 [0.45, 0.52] & 0.0038 [-0.010, 0.064]\\
\hline
matrix weights & $w_\text{GRM}=0.1$, $w_\epsilon=0.9$ & $w_\text{GXM}=1$, $w_\epsilon=0$ & $w_\text{MXM}=0.4$, $w_\epsilon=0.6$ & $w_\text{GXM}=0.8$, $w_\text{MXM}=0.1$, $w_\epsilon=0.1$ & NA\\
\hline
\end{tabular}
\begin{flushleft}GRM = genetic relationship matrix, GXM = gene expression correlation matrix, MXM = microRNA expression matrix, baseline = multivariate prediction model including top genes, top microRNAs, and first ten principal components
\end{flushleft}
\label{tab:igrowth}
\end{table}

\begin{table}[!ht]
\caption{
\bf{OmicKriging and other model performance for prediction of statin-induced change in LDL cholesterol levels.}}
\begin{tabular}{|p{1.5cm}|p{2cm}|p{2cm}|p{2cm}|p{2cm}|p{2cm}|p{2cm}|}
\hline
Model: & OmicKriging Hopewell GRM (n=562) & OmicKriging Hopewell GRM (n=461) & OmicKriging RHOA GXM (n=461) & OmicKriging Hopewell GRM + RHOA GXM (n=461) & Polyscore top1K (n=562) & baseline (n=461)\\
\hline
mean R$^2$ [95\% CI] & 0.037 [0.026, 0.049] & 0.024 [0.015, 0.035] & 0.0021 [-0.00086, 0.0057] & 0.025 [0.014, 0.036] & 0.016 [0.0048, 0.030] & -0.00014 [-0.0022, 0.0070]\\
\hline
matrix weights & $w_\text{GRM}=0.8$, $w_\epsilon=0.2$ & $w_\text{GRM}=0.9$, $w_\epsilon=0.1$ & $w_\text{GXM}=0.1$, $w_\epsilon=0.9$ & $w_\text{GRM}=0.6$, $w_\text{GXM}=0.3$, $w_\epsilon=0.1$ & NA & NA\\
\hline
\end{tabular}
\begin{flushleft}Hopewell GRM = genetic relationship matrix of Hopewell et al. \cite{hopewell2013impact} 50kb SNPs, RHOA GXM = gene expression correlation matrix of RHOA regulatory pathway, Polyscore top1K = polygenic score model of top 1K SNPs, baseline = multivariate prediction model including top genes, the 45 Hopewell et al. SNPs, and first ten principal components
\end{flushleft}
\label{tab:dldlc}
\end{table}

\begin{table}[!ht]
\caption{
\bf{OmicKriging and other model performance for prediction of disease risk in the WTCCC.}}
\begin{tabular}{|c|p{2.2cm}|p{2.2cm}|p{2.2cm}|p{2cm}|p{2cm}|p{2cm}|}
\hline
Disease & Results & OmicKriging Single GRM & OmicKriging Double GRM  &  BSLMM (SD)* &  baseline\\
\hline
BD & mean AUC [95\% CI] & 0.666 [0.661, 0.670] & 0.665 [0.660, 0.670] &  0.65 (0.02) &  0.594 [0.590,0.597]\\
\hline
 & matrix weights & $w_\text{GRM}=0.3$, $w_\epsilon=0.7$ & $w_\text{GRM}=0.4$, $w_\text{GRMk}=0$, $w_\epsilon=0.6$ & NA  & NA\\
\hline
CAD & mean AUC [95\% CI] & 0.598 [0.593, 0.604] & 0.610 [0.605, 0.616] &  0.60 (0.03) &  0.575 [0.572, 0.578]\\
\hline
 & matrix weights & $w_\text{GRM}=0.4$, $w_\epsilon=0.6$ & $w_\text{GRM}=0.6$, $w_\text{GRMk}=0.1$, $w_\epsilon=0.3$ & NA  & NA\\
\hline
CD & mean AUC [95\% CI] & 0.641 [0.636, 0.646] & 0.707 [0.704, 0.710] &  0.68 (0.02)  & 0.669 [0.667, 0.670]\\
\hline
 & matrix weights & $w_\text{GRM}=0.3$, $w_\epsilon=0.7$ & $w_\text{GRM}=0.4$, $w_\text{GRMk}=0.1$, $w_\epsilon=0.5$ & NA & NA\\
\hline
HT & mean AUC [95\% CI] & 0.611 [0.606, 0.616] & 0.611 [0.605, 0.616] &  0.60 (0.02) &  0.493 [0.485, 0.499]\\
\hline
 & matrix weights & $w_\text{GRM}=0.3$, $w_\epsilon=0.7$ & $w_\text{GRM}=0.3$, $w_\text{GRMk}=0$, $w_\epsilon=0.7$ & NA & NA\\
\hline
RA & mean AUC [95\% CI] & 0.649 [0.644,0.654] & 0.747 [0.744, 0.751] &  0.72 (0.01) & 0.668 [0.666, 0.669]\\
\hline
 & matrix weights & $w_\text{GRM}=0.2$, $w_\epsilon=0.8$ & $w_\text{GRM}=0.2$, $w_\text{GRMk}=0.5$, $w_\epsilon=0.3$ & NA  & NA\\
\hline
T1D & mean AUC [95\% CI] & 0.713 [0.709, 0.717] & 0.891 [0.889, 0.892]  & 0.88 (0.01) &  0.772 [0.770, 0.773]\\
\hline
 & matrix weights & $w_\text{GRM}=0.4$, $w_\epsilon=0.6$ & $w_\text{GRM}=0.3$, $w_\text{GRMk}=0.4$, $w_\epsilon=0.3$ & NA  & NA\\
\hline
T2D & mean AUC [95\% CI] & 0.618 [0.613, 0.624] & 0.634 [0.630, 0.639] &  0.61 (0.03) & 0.609 [0.607, 0.611]\\
\hline
 & matrix weights & $w_\text{GRM}=0.3$, $w_\epsilon=0.7$ & $w_\text{GRM}=0.7$, $w_\text{GRMk}=0.1$, $w_\epsilon=0.2$ & NA & NA\\
\hline
\end{tabular}
\begin{flushleft}*values reported in Zhou et al.\cite{zhou2013polygenic}, BSLMM = Bayesian sparse linear mixed model, SD = standard deviation, baseline = polygenic score model of known GWAS SNPs and first ten principal components, AUC = area under the receiver operating characteristic curve, CI = confidence interval, GRM = genetic relationship matrix (all SNPs), GRMk = genetic relationship matrix of known GWAS SNPs, BD=bipolar disorder, CAD=coronary artery disease, CD=Crohn's disease, HT=hypertension, RA=rheumatoid arthritis, T1D=type 1 diabetes, T2D=type 2 diabetes \\
\end{flushleft}
\label{tab:disease}
\end{table}

\end{document}